# Review of recent results on streamer discharges and discussion of their relevance for sprites and lightning

Ute Ebert[1,2], Sander Nijdam[2], Chao Li[1], Alejandro Luque[1,3], Tanja Briels[2,4], Eddie van Veldhuizen[2]

1. Centrum Wiskunde & Informatica (CWI), Amsterdam, The Netherlands
2. Faculty of Physics, Eindhoven University of Technology, Eindhoven, The Netherlands
3. Instituto de Astrofísica de Andalucía (IAA), CSIC, Granada, Spain
4. Department of Engineering Physics, Fontys University of Applied Sciences, Eindhoven, The Netherlands

**Abstract**.

It is by now well understood that large sprite discharges at the low air densities of the mesosphere are physically similar to small streamer discharges in air at standard temperature and pressure. This similarity is based on Townsend scaling with air density. First the theoretical basis of Townsend scaling and a list of six possible corrections to scaling are discussed; then the experimental evidence for the similarity between streamers and sprites is reviewed. We then discuss how far present sprite and streamer theory has been developed, and we show how streamer experiments can be interpreted as sprite simulations. We review those results of recent streamer research that are relevant for sprites and other forms of atmospheric electricity and discuss their implications for sprite understanding. These include the large range of streamer diameters and velocities and the overall 3D morphology with branching, interaction and reconnection, the dependence on voltage and polarity, the electron energies in the streamer head and the consecutive chemical efficiency and hard radiation. New theoretical and experimental results concern measurements of streamer spectra in air, the density dependence of streamer heating (hot leaders are unlikely at 80 km altitude and cold streamers are unlikely in liquids), and a discussion of the influence of magnetic fields on thermal electrons or on energetic electrons in streamers or sprites.

# 1. Introduction

## 1.1 Mechanisms and scales in sprite discharges

A sprite discharge is a physical process that involves exceptionally many length scales: its emergence and evolution depends on the ionization and density profiles [Barrington-Leigh *et al.*, 2001; Hiraki and Fukunishi, 2007] in mesosphere and lower ionosphere (roughly at 40 to over 100 km altitude [Sentman *et al.*, 1995]) and on the evolution of lightning currents between cloud and ground at 0 km altitude [Pasko, 2006; Williams *et al.*, 2007; Hu *et al.*, 2007]. Sprite discharges can consist of (ten-)thousands of growing channels with diameters of the order of tens to hundreds of meters [Gerken *et al.*, 2000]. They must have an inner structure with space charge layers that can have widths of the order of meters or less, as theory evidences. (References to theory will be given below.) To resolve many features of these structures, a density approximation for electrons and ions is sufficient, but for a detailed understanding, e.g., of the possibility that electrons run away from the streamer tip and create X-rays and gamma-rays, the dynamics of individual electrons within the discharge channel has to be understood.

## 1.2 The state of theory and simulations

Obviously, it is nearly impossible to catch all features from the earth-ionophere scale down to the mean free path length of individual electrons within one theory or simulation program though this is desirable not only for understanding discharge processes in the terrestrial atmosphere, but also for extrapolating to atmospheric discharges on other planets [Yair et al., 2009]. Theory in the past years has succeeded in establishing various partial results, relating two or more phenomena on different scales.

First, there is an increasing understanding on which charge moment changes of the thundercloud-earth system and which ionization profiles of the upper mesosphere and lower ionosphere are required to raise the electric field to values above the breakdown field in the lower ionosphere or upper mesosphere [Pasko et al., 1996, 1997, 1998; Raizer et al., 1998; Barrington-Leigh et al., 2001; Hu et al., 2002, 2007; Hiraki & Fukunishi, 2007; Adachi et al., 2008; Li J et al., 2008]. These models include the parent lightning stroke as well as the ionization and density profiles of upper meso- and lower ionosphere and identify a necessary condition for sprite generation. Exceeding the breakdown field and creating a sufficiently sharp ionization profile at the lower edge of the ionospheric E region is a necessary, but not a sufficient criterion for sprite emergence, and these studies do not resolve whether the sprite really does emerge and what its properties are.

Second, the propagation of single discharge streamers was simulated by many authors in numerous papers since the seminal work of Dhali & Williams [1985, 1987]. An overview would include many tens of articles by authors as Wu, Kunhardt, Vitello, Penetrante, Bardsley, Babeva, Naidis, Kulikovskii, Starikovskii, Pancheshnyi, Bourdon and others and will not be attempted here. Single sprite streamers were analyzed in particular by Liu and Pasko [2004, 2006] and Liu et al. [2009]. The scaling of streamers with gas density was also tested in simulations by Pancheshnyi et al. [2005], by Pasko [2007] and by Luque et al. [2007, 2008b]. These models resolve the inner structure of a single discharge channel in a density approximation for electrons and ions within a given electric field and for constant air density; how this field is generated by the effect of the lightning stroke on the ionization profiles in meso- and ionosphere is outside the scope of these models. Several streamers and their interactions up to now have been studied in only three papers [Naidis, 1996; Luque et al., 2008a, 2008b]; they are discussed in section 3.3.1.

The first and the second problem are joined in a recent simulation study of Luque & Ebert [2009] where the numerical grid is refined adaptively; in this case study the electrical current of the lightning stroke, the lower edge of the ionosphere and the emerging halo and sprite are resolved, and the change of air density with altitude is taken into account. The emergence of a single sprite channel from the ionosphere is simulated up to the moment when it breaks up into many channels.

Third, the possible run-away of individual electrons from a streamer or a sprite requires resolving the dynamics of single electrons in the high field region at the streamer tip. Recent progress in the development of simulation methods and in the evaluation of results by Moss et al. [2006], Li C et al. [2007, 2008a, 2008b, 2009, 2010] and Chanrion & Neubert [2008, 2009] is discussed in section 3.4.1.

Fourth, to proceed to a proper description of a whole branched tree of streamers, the microscopic models have to be reduced appropriately. A characterization of complete streamer heads by charge, radius, voltage, enhanced field, velocity etc. has been proposed by Luque et al. [2008c]. Naidis [2009] suggests a relation between velocity, radius and enhanced field. Along these lines a new generation of dielectric breakdown models can be developed that contain more microscopic input than current dielectric breakdown (DBM) models based on the concepts of Niemeyer et al. [1984].

**1.3 A joint approach by theoretical and experimental simulations**

From what is said above, it is clear that sprite simulations on the computer are challenging and time consuming, and they are still constrained to one or two streamers, or to infinitely many under simplifying assumptions. Other approaches to attack this problem are therefore extremely valuable. Only a combination of results derived with different methods will lead to understanding; in this spirit the Leiden workshops in 2005 and 2007 were organized, and presentations are collected in a cluster issue on streamers, sprites and lightning in J. Phys. D [Ebert & Sentman, 2008]. In the present manuscript, we in particular will elaborate the question how far sprites can be simulated in laboratory experiments on streamers while also touching on streamer simulations where appropriate.

The paper is organized as follows. In section 2, first the theoretical basis of the similarity between streamers and sprites is discussed and then possible sources of corrections; such sources can be intrinsic in the discharge, from boundaries, from stochastic fluctuations, due to density gradients or external ionization sources, from magnetic fields or due to intrinsic heating. Then the similarity is confirmed by comparing particular experimental streamer results with sprite observations. On this basis we review recent streamer results in section 3 and discuss their relation to sprite observations and possible implication for sprites. In particular, we discuss morphology, diameters, velocities and currents and their dependence on voltage and polarity in 3.1; and we present spectra in 3.2, the 3D structure of streamer trees including branching and interaction of streamers in 3.3, and finally streamers as chemical reactors and electron accelerators, and the X-ray and γ-ray production in 3.4.

## 2. Similarity of streamers in gases of different density

### 2.1 The theoretical basis of similarity and its corrections

*2.1.1 Similarity laws or Townsend scaling*

Streamer discharges in gases with the same composition but with different densities (as is the case for atmospheric air up to 90 km altitude) can be physically similar. Physical similarity means that the phenomena are the same if lengths, times, fields, densities of charged or excited particles etc. are measured on appropriate density dependent scales; we will refer to the precise similarity laws as to *Townsend scaling*, and we will discuss corrections to the scaling laws in the next subsections.

The physical basis of Townsend scaling is the following. The length scale of the discharge processes in the streamer tip is determined by the mean free path length $\ell_{MFP}$ of an electron between collisions with the neutral gas molecules, if the electron density is so low that collisions with neutral molecules dominate over collisions with ions or other electrons. (Below we will come back to this constraint on the relative ionization density that is inherent in present streamer and sprite models.) If the electrons predominantly collide with neutral molecules, the mean free path length of the electrons is inversely proportional to the molecule number density $n$ of the gas: $\ell_{MFP} \sim n^{-1}$. Therefore all lengths determined by electron motion scale like $n^{-1}$. The kinetic energies $E$ of the electrons have to reach the ionization threshold of the gas molecules which is a molecule specific value and independent of the gas density; therefore the characteristic electron energies do not depend on gas density: $E \sim n^0 = 1$. As electron energies and velocities are related through $E = \frac{1}{2} mv^2$, the characteristic electron velocities are independent of gas density n as well: $v \sim n^0$. As lengths scale like $1/n$ and electron velocities are independent of $n$, times have to scale as $1/n$ as well $t \sim 1/n$. As energies are independent of $n$, voltages are independent of $n$: $U \sim n^0$, and therefore electric fields scale as $n$: $\mathbf{E} \sim n$, as they have dimension of voltage over length. To take this scaling into account, the unity Townsend or Td has been introduced in gas discharge physics for the reduced electric field strength $|\mathbf{E}|/n$: 1 Td = $10^{-17}$ V cm$^2$. Without doing extensive historical research, it can be stated that these relations were certainly known by German, Dutch and English scientists before World War II. Originally, not the density $n$ was used, but the pressure $p$ at room temperature; the two quantities are related in very good approximation by the ideal gas law $p = n\, k_B T$.

It was argued above that the characteristic lengths and times of electron motion scale with the inverse gas density $1/n$. But during the rapid evolution of ionization avalanches or of streamer ionization fronts, ions and excited species are just passively created by a rapidly passing electron distribution and essentially don't move during the passage of the front; therefore all length and time scales of the discharge scale with the inverse gas density $1/n$. This can be seen more clearly by performing a dimensional analysis of the discharge equations as is done, e.g., in [Ebert et al., 2006]. The similarity means that a decrease of gas density acts as magnifying glass and slow motion player.

The scaling relations above result from the dynamics of single electrons in a given field; therefore they apply to avalanches as well as to streamers. On the other hand, the ionization density inside the streamer results from a dimensional analysis of the Poisson equation that creates the nonlinearity: the charge density integrated over the width of the ionization front has to screen the electric field. As fields scale like $n$ and lengths scale like $1/n$, densities of charged particles scale as $n_e \sim n^2$ [Pasko et al., 1998, 2006, Rocco et al., 2002; Ebert et al., 2006]; this means that the relative electron density $n_e/n$ scales like $n$ while the total number of electrons in a section of a streamer scales like the total electron density times the relevant volume $n^2/n^3 = 1/n$. Therefore streamers at mesospheric altitudes have a lower relative electron density and a larger total number of electrons than at sea level. This has two direct consequences. First, approximating streamer dynamics by densities of non-interacting electrons works better at lower gas densities, because electron-electron interaction is less important, and because the larger total number of electrons leads to better statistics. Second, the total time integrated luminosity of a streamer or sprite head scales with inverse gas density $1/n$, or in other words, a similar streamer or sprite at lower density emits more light.

As velocities don't scale with density in similar streamers, electric current densities scale like electron densities $n_e \sim j \sim n^2$, and the electric current integrated across the whole streamer channel does not depend on $n$: $I \sim n^0$.

The essentials of the similarity relations are summarized as follows. Streamers in the same gas at different density $n$ are similar, if the same voltage $U$ is applied in a geometry whose lengths scale as *1/n*. All length and time scales of the discharge will then scale as *1/n*, while the velocities *v* and the electric currents integrated over the streamer cross-sections *I* are independent of *n*. The relative electron density $n_e/n$ scales as *n*. The absolute electron density $n_e$ and the time integrated light emission density from the streamer head scale as $n^2$, and the time integrated light emission from the complete head scales as *1/n*.

These similarity relations should not be misunderstood in the sense that all streamers are similar. In fact, recent investigations of streamers at fixed gas composition and density have shown that diameters and velocity can vary largely, mainly dependent on the applied voltage, as we will elaborate in section 3. We suggest that there is the same wide variation of diameters and velocities in sprites at fixed altitudes as we find in streamers at standard temperature and pressure.

But first we will focus on the theoretical range of validity of the similarity laws and on their experimental tests in the remainder of section 2. The similarity laws apply in particular to the fast processes at the growing tip of a streamer; they are dominated by the collisions of electrons with neutral gas molecules. Corrections to scaling come from different sources.

*2.1.2 Corrections to Townsend scaling due to gas discharge processes*

*All processes that involve the interaction of two charged particles or of two neutral particles or processes evolving in two or more steps will in general break the similarity laws.* The most important example is the quenching of excited nitrogen molecules by collision with other molecules which suppresses the photo-ionization rate at pressures above 80 mbar at standard temperature [Teich, 1967; Zhelezniak et al., 1982] (i.e., roughly below 18 km altitude in the atmosphere); photoionization is generally assumed to explain the propagation of positive streamers in air, and its quenching is relevant for positive streamer heads in air at standard temperature and pressure (STP, i.e., 20ºC., 1 bar). Other processes that break the scaling laws are electron-ion recombination and three-body attachment that decrease the conductivity inside the streamer channel after its generation [Pasko, 2007]; therefore rescaled streamers whose head dynamics is

similar, lose the conductivity inside the generated channel faster at higher gas densities, even on their intrinsically faster time scale.

An important point that has found little attention is the heating of the streamer channel: the electric currents inside the streamer channel dissipate a power density that scales like $j\,E \sim n^3$ and the dissipated power per neutral molecule inside the streamer channel therefore scales as $n^2$. Taking the scaling of streamer time scales with density into account, the dissipated electrical energy per gas molecule scales as $n$ in similar streamers over similar times, as long as the streamer maintains its conductivity. We will elaborate the consequences of this observation in section 2.1.7.

*2.1.3 Corrections to Townsend scaling due to electrodes or other material boundaries*

*In most streamer experiments, the streamers start from an electrode. Shape and surface properties of the electrodes cannot be rescaled with changing gas density; the same holds for dust particles, droplets or other material boundaries.* Therefore the similarity laws do not hold in the immediate neighborhood of electrodes as discussed and illustrated by Briels et al. [2008b]; the role of a needle electrode during streamer inception at different densities is illustrated in the photographs by Briels et al. [2008c]. This is why Liu et al. [2006] in their simulations rescaled the electrode radius with air density. But this is not easily done in experiments, and furthermore the electrode processes might be governed by microscopic roughness that can not be rescaled anyhow. However, Fig. 1 illustrates that the morphologies of propagating streamers resemble each other for different electrode configurations and for different distances from the electrodes if they only have propagated about twice their diameter from the electrode; therefore it is justified to compare these streamers with sprites that do not start from electrodes. In fact, the left panel in Fig. 1 illustrates that a thick and well conducting streamer or sprite can generate similarly branched trees of thinner streamers as a needle electrode.

*2.1.4 Corrections to Townsend scaling due to stochastic effects*

*As said above, the total number of electrons in a streamer head scales with gas density as 1/n which is important for all stochastic effects.* When streamer formation requires approximately $10^8$ electrons at standard temperature and pressure according to

the Raether-Meek criterion (as reinvestigated recently by Montijn et al. [2006a]), then it requires approximately $10^{13}$ electrons at 83 km altitude. (Here and later, we use the approximation $n(z) = n_0 \exp[-z/h]$, $h = 7.2$ km, $n_0 = 2.5 \cdot 10^{19}$ cm$^{-3}$, for the air density $n$ as a function of altitude $z$, which is a very good approximation for the ISO standard atmosphere [ISO, 1975].)

It has been shown [Arrayas et al., 2002; Rocco et al., 2002; Montijn et al., 2006c] that branching of streamers is a nonlinear bifurcation process that can occur without any fluctuations. However, as discussed by Ebert et al. [2006], electron density fluctuations can accelerate the branching process. Due to the smaller total electron number in the streamer, these fluctuations are stronger at higher gas densities in otherwise similar streamers. These higher fluctuation rates would make a streamer branch earlier at higher gas densities; this branching enhancement mechanism differs from the mechanism by suppressed photoionization suggested by Liu and Pasko [2006]. Finally, rare events like the run-away of electrons from a streamer when the run-away condition is not yet reached, are more likely with larger electron number, i.e., with lower gas density. Accurate inclusion of such stochastic effects into streamer simulations requires following the single electron dynamics at the very tip of the streamer through a Monte Carlo procedure. Appropriate numerical methods are recently being developed by Li C et al. [2007, 2008a, 2008b, 2009, 2010] and by Chanrion & Neubert [2008, 2009]. (We will return to these papers in 3.4.1.)

*2.1.5 Corrections to Townsend scaling due to ionization sources or density changes*

Similarity can also be broken by external ionization sources that create background ionization; furthermore, streamer experiments cannot reproduce the profiles of gas and ionization densities varying with atmospheric altitude. We note that in the sprite simulation by Luque & Ebert [2009] that takes these height profiles into account, the effect of the changing air density is not visible in the sprite. This is, because the sprite branched already after 2 km of propagation where the density varies by only 30%, and the further evolution was not followed.

*2.1.6 Corrections to Townsend scaling due to the geomagnetic field*

As thermal electrons are magnetized in the geomagnetic field at mesospheric altitudes, sprite discharges could be magnetized as well, and we here analyze this situation. Based on the data presented in Fig. 2, we conclude that thermal electrons indeed can be magnetized above 50 to 70 km, while the energetic electrons in a streamer or sprite ionization front will hardly be influenced by the geomagnetic field even at 90 km altitude. This statement stems from comparing the cyclotron frequency in the geomagnetic field with the collision frequency of electrons [Egeland et al., 1973]. The collision frequency depends on air density and on electron energy.

The cyclotron frequency is $v_c = qB/(2\pi m)$, where $q$ is elementary charge, $B$ is the magnetic field and $m$ is electron mass. The cyclotron frequency in a geomagnetic field of 30 μTesla, i.e., close to the equator, is $v_c = 0.8$ MHz.

The electron-neutral collision frequency $v$ is calculated as a function of mean electron energy $\epsilon$ at different altitudes $z$ as

$$\nu(\epsilon, z) = \sqrt{\frac{2\epsilon}{m}} \sum_{i=1}^{N_g} \sum_{j=1}^{M_i} \sigma_{ij}(\epsilon) n_i(z)$$

where $N_g$ is the number of different gas species labeled by *i*, and $M_i$ is the number of collision processes for the *i*th gas species. The total cross section is calculated by summing the specific collision cross sections $\sigma_{ij}(\epsilon)$ over all collision processes of a given gas species ($N_2$, $O_2$ and Ar in air) multiplied with their partial density $n_i(z)$. All partial densities scale as the total density: $n_i(z) \sim n(z)$; therefore the collision frequency scales with $n$ as well: $v(\epsilon,z) = v(\epsilon,0)\, n(z)/n_0$, where $n_0$ is air density at standard temperature and pressure (STP). We included 44 collision processes (25 for $N_2$, 16 for $O_2$ and 3 for Ar) using the electron-neutral cross sections from the Siglo database http://www.siglo-kinema.com [Morgan et al., 1995]. The collision frequencies plotted as dotted lines in Fig. 2 correspond to air densities of $n/n_0$ = 1, 0.1, 0.01, 0.001, 0.0001, 0.00001. They roughly correspond to altitudes of 0, 17, 33, 50, 66, and 83 km with the approximation $n = n_0\, exp[-z/h]$, $h = 7.2$ km.

The electron energy in Fig. 2 ranges from 0.01 to 1 000 eV, where 0.01 eV is the thermal energy at a temperature of 116 K. The collision frequency $v$ in general increases with electron energy up to ~100 eV, with some intermediate peak at 2-4 eV due to the rotational and vibrational collisions. Beyond ~150 eV, the collision frequency decreases again (creating the possibility of electron run-away).

The red solid line in Fig. 2 relates the reduced electric field (right axis) to the mean electron energy (bottom axis). The relation is obtained from electron swarm simulations performed as in [Li C et al., 2007, 2010] after the swarm has equilibrated to a uniform electric field (which occurs for reduced field strengths below 260-300 kV/cm according to Phelps et al. [1987] and Kunhardt and Zeng [1988]; above that value the whole swarm starts running away).

The reduced electric field $|\mathbf{E}|n_0/n$ at the streamer head has to be higher than the breakdown field of 32 kV/cm; in Fig. 2 we indicate typical reduced fields at the streamer head of 50, 100 and 250 kV/cm and their mean electron energies. Clearly the collision frequency at these energies is about two orders of magnitude larger than the cyclotron frequency even at 83 km altitude and increases strongly at lower altitudes. We conclude that the geomagnetic field is essentially negligible for sprites, in accordance with observations. Experimental results supporting this statement will be presented in section 2.3.3.

*2.1.7 Possible non-existence of streamers in liquids and of leaders in the mesosphere*

It was stated above that the electron density in similar streamers scales as $n^2$ with the neutral gas density $n$, and that the relative electron density therefore scales as $n_e/n \sim n$. If the relative electron density $n_e/n$ of a streamer in room air is of the order of $10^{-5}$ [Ebert et al., 2006], then it is about $10^{-10}$ in sprites at 80 km altitude, and it is of the order of $10^{-2}$ at liquid densities. As said above, the dissipated electrical energy per gas molecule scales as $n$ in similar streamers as well. There are recent indications (oral communication with Z Liu, AJM Pemen and EM van Veldhuizen in Eindhoven) that very strong and fast voltage pulses can heat a streamer in normal room air within 10 ns. Very fast heating could also explain the commonly observed bubble formation for streamers in water and other dielectric liquids [Kolb et al., 2008]. We here formulate the hypothesis that due to the strong heating proportional to the medium density $n$, cold streamers at liquid densities do not really exist, but rather immediately form a type of hot leader state – apart from the question whether the streamer approximation of electrons colliding predominantly with neutral gas molecules is still justified. Note that our concept of fast immediate streamer heating at high densities differs from the concept suggested by Tardiveau et al. [2001] and Marode et al. [2009] who argued that heat diffusion is suppressed at higher densities, but did not consider heat production.

On the other end of the density scale, at sprite altitudes, ionization and Ohmic heating is weak and streamer conductivity is maintained by low rates of electron attachment and electron-ion recombination. Therefore we formulate the hypothesis that a transition to a hot leader state and later to a spark can hardly occur in the mesosphere.

**2.2 Experimental confirmation of similarity**

As argued above, there might be corrections to the similarity laws, in particular, for pressures above ~80 mbar at room temperature, when photo-ionization is increasingly quenched (cf. section 2.1.2). But simulations (Liu & Pasko [2004, 2006], Pasko [2007], Luque et al. [2007, 2008b]) show that the effect is minor as long as the electric field far ahead of the streamer is below the breakdown threshold. The similarity of experimental streamers and observed sprites is supported by the following observations.

*2.2.1 Morphology, minimal diameters and velocities*

The high speed video observations of sprites by Cummer et al. [2006] shows how a single sprite streamer emerges out of a halo, shoots downwards and breaks up into many branches. The break-up into branches has a similar morphology as the streamers in Fig. 1.

Photographs of streamers with high spatial resolution as in Fig. 1 show that the diameters of the streamers vary largely (as we will discuss in more detail in 3.1.1), but that there seems to be a minimal diameter. That there should be a minimal streamer diameter is supported by the following consideration [Briels et al., 2008b]: the field enhancement at the streamer head is created by a thin space charge layer. This space charge layer has a minimal width, given by the inverse of the maximum of the Townsend ionization coefficient (this coefficient is the product of the cross section of one molecule times the particle number density of the molecules). The streamer diameter needs to be larger than the width of the space charge layer for field enhancement to be efficient. Therefore there is a minimal streamer diameter.

The minimal streamer diameter is an appropriate quantity to test the similarity relations discussed in section 2.1.1: the diameter should scale as the inverse gas density *1/n*. This means that the reduced diameter, i.e., the product of diameter and density, should be independent of density. This relation was already tested over almost two decades of density by Briels et al. [2008b]; here we present improved measurements in Fig. 3. The main improvement was a new lens system allowing zooming in closer into the discharge so that artifacts due to cross-talk between camera pixels (as described by Briels et al. [2006]) could be further reduced, and furthermore the evaluation procedure of the measurements was improved, for details we refer to Nijdam et al. [2010]. Fig. 3 shows that the reduced diameter is nearly constant for pressures between 25 and 200 mbar at room temperature, both for artificial air (a mixture of 80% $N_2$ and 20% $O_2$) and for pure nitrogen (with less than 1 p.p.m. impurities). The increase of reduced diameter for pressures above 200 mbar could either be due to corrections to the similarity laws (though one would expect corrections in the opposite direction) or because streamers again become so thin that they are not sufficiently resolved by the camera. As in [Briels et al., 2008b], the smallest reduced diameter of the telescopic sprite measurements of Gerken et al. [2000] is plotted in Fig. 2 as well; the error bar for Gerken's result accounts for the error in diameter and in altitude given in her paper. Though both the true streamer diameter and the density vary by 5 orders of magnitude within the plot, the reduced minimal diameters in air agree in a linear plot within the error bar. It should be noted that measurements allowing the determination of the sprite diameter were essentially only done by Gerken et al. [2000]. Therefore it is very well possible that these sprite streamers did not have minimal diameter and therefore have a larger reduced diameter than the minimal streamers in our measurements. We will give more arguments hinting into this direction in 3.1.1. More observational work on diameters of sprite streamers would be very desirable.

A related observation is that the minimal velocities of streamers and sprites agree as well; they are predicted to be independent of density according to section 2.1.1. While both for streamers and for sprites, a large range of velocities can be found in the literature, a minimal velocity of ~$10^5$ m/s is found for laboratory streamers by Briels et al. [2006, 2008b] and the same value was found for tendrils in sprites by Moudry et al. [2002].

*2.2.2 Light emission focused at the streamer tips*

Another strong indication for the physical similarity of streamers and sprites is their optical signature at very short exposure times. When intensified CCD cameras with exposure times as low as 30 or 5 ns became available, it was recognized by Blom et al. [1994, 1997] that streamers in air emit light essentially only at their growing tips. The effect is illustrated in a didactical manner by Ebert et al. [2006] and also in the right column of Fig. 1 in the present manuscript; it is based on the fact that the light is predominantly emitted by the excited state $N_2$ ($C^3\Pi_u$) that has a life time of only ~1 ns in air at standard temperature and pressure (STP air). Other illustrative photographs were taken by Pancheshnyi et al. [2005] with a stroboscopic camera with an exposure time of 1.3 ns and a repetition rate of 1/(5ns). Nudnova & Starikovskii [2008] even reconstruct the cap-formed layer of instantaneous emission around the streamer head from ICCD photographs with 200 ps exposure.

If streamers and sprites are physically similar, the same light emitting tips should be present in sprites. Indeed they were found by Stenbaek-Nielsen et al. [2007] and McHarg et al. [2007] as discussed in more detail by Stenbaek-Nielsen & McHarg [2008].

*2.2.3 Streamers or sprites in strong magnetic fields*

In section 2.1.6, we presented a theoretical argument that the geomagnetic field should not have a visible effect on sprites. Laboratory experiments on streamers in high magnetic fields have been performed in the High Magnetic Field Lab in Nijmegen by Manders et al. [2008]. In their experiments on streamers at 200 to 600 Torr in fields up to 12.5 Tesla, they find Hall angles up to 10°. Extrapolating 12.5 Tesla at 600 Torr to 83 km altitude yields 160 µTesla (as effects of magnetic fields scale as gas density *n*, just like those of the electric fields). Therefore a magnetic field of 160 µTesla would generate a Hall-angle of 10° at 83 km altitude, while the geomagnetic field close to the equator is less than 1/5 of that value. We conclude that the extrapolation of these laboratory measurements indicate as well that the geomagnetic field should not influence sprite propagation in any visible manner; this is in agreement with actual observations of sprites where such an effect never was found.

## 3. Streamer experiments as sprite simulations

Having established the approximate similarity of streamers and sprites, we now review recent streamer experiments and emphasize those results that are important for the interpretation of sprite observations. Given the limitations of current streamer and sprite simulations discussed in section 1.2, we suggest that streamer experiments are an important complementary tool to simulate and understand sprites.

**3.1 Morphology, diameters, velocities and currents, voltage and polarity dependence**
*3.1.1 Morphology and diameters of positive streamers as a function of voltage*

Recently developed voltage sources can raise the electric field to values much above the breakdown value within tens of nanoseconds; they can create discharge trees where individual streamers have a large variety of diameters and velocities. Examples of morphology in the case of a needle-to-plate electrode geometry are shown in Fig. 1. As described by Briels et al. [2006, 2008a], both the diameter and the velocity of the streamers emitted from a needle electrode can vary by one or two orders of magnitude depending on the applied voltage, as illustrated in Fig. 1. The thick and fast streamers branch into thinner and slower streamers; and the process continues until the minimal streamer diameter is reached; streamers of minimal diameter do not branch anymore. In the case of a wire electrode where the electric field only decays like $1/r$ with distance r (while it decays like $1/r^2$ ahead of a needle), Winands et al. [2008] also have observed the streamer diameter and velocity to increase in time, similarly to what has been observed in sprites [Li J & Cummer, 2009; Liu et al., 2009]. We note that Li J & Cummer [2009] argue that an inhomogeneous field as in the experiments would not occur in sprites; however, the destabilization of the lower edge of the E region of the ionosphere through a screening ionization wave as described by Luque & Ebert [2009] can generate an inhomogeneous field, and the field becomes more inhomogeneous through the presence of the streamer.

The streamer diameters in STP air measured by Briels et al. [2008a] for voltages of 5 to 96 kV range from 0.2 to 3 mm, their velocities from $10^5$ to $4 \cdot 10^6$ m/s (where we recall that velocities and voltages don't scale with density). These changes of diameter and velocity by one to two orders of magnitude are illustrated in Fig. 1. When air density decreases by 5 orders of magnitude (as from 0 to 83 km altitude with the approximation n ~ $e^{-z/h}$ where z is altitude and h = 7.2 km), lengths increase by 5 orders of magnitude, i.e., 1 cm becomes 1 km, i.e., sprite streamers at 83 km altitude similar to Briels' laboratory streamers would have diameters of 20 to 300 m. But the measurements of Nijdam et al. [2010] have further decreased the minimal streamer diameter in STP air to 0.12 mm; according to the measurements presented in Fig. 3, the minimal sprite diameter at 83 km altitude is 12 m (or 1.2 m at 66 km altitude). The error bar at the sprite diameter in Fig. 3 indicates that 12 m is below the resolution of Gerken et al. [2000]. Furthermore, as Fig. 1 illustrates, such minimal streamers would be very dim and not propagate far, and they would not branch while propagating downwards; we therefore suggest that all observed sprite streamers have a diameter larger than minimal (as already suggested by Fig. 3). This suggestion agrees with the observation that the streamer velocity is larger than $10^5$ m/s. Indeed, Briels et al. [2008a] have derived a completely empirical fit formula to their experimental data for the relation between diameter and velocity that is reasonably confirmed by simulations by Luque et al. [2008c]; when similarity laws are introduced, this relation between velocity *v*, diameter *d* and air density *n* is

$v = 5 \cdot 10^5$ m/s $[(d\ n)/(n_0\ \text{mm})]^2$,

where $n_0$ is air density at sea level. (Naidis [2009] suggests an analytical argument for a similar relation that is based on a number of assumptions.) Briels' relation suggests that a sprite streamer at 83 km altitude would have a velocity of $10^7$ m/s, if its diameter is ~400 m.

We note that the upper limit of streamer diameters and velocities in our experiments is set by the available voltage sources, and that we are currently working on improvement. A streamer powered by a voltage of the order of MV as available in and around thunderclouds could be much faster and thicker than the streamers investigated by Briels et al. [2008a] and Winands et al. [2008] that are powered by up to 96 kV with a minimal voltage rise time of 15 ns. We stress that streamers with small diameters and velocities are the easiest to generate in the lab, and the most difficult to observe in sprites. For large diameters and velocities, the situation is the reverse.

*3.1.2 Currents and polarity dependence*

The electric currents measured by Briels et al. [2006] vary from ~10 mA to 25 A per streamer channel with increasing streamer diameter. We recall that currents don't scale with density; therefore similar currents should flow in sprites.

Furthermore, until here only positive streamers were discussed. For negative streamers, Briels et al. [2008a] find that they largely resemble diameters and velocities of positive streamers for voltages above ~40 kV, but they do not reach the propagation lengths of positive streamers. For lower voltages, negative streamers are difficult to initiate. Luque et al. [2008c] suggest that this is because the space charge layer at the head of a positive streamer is formed by relatively immobile ions and a depletion of electrons while in negative streamers it is formed by an overshoot of electrons. These electrons in the negative streamer head can drift away in the electric field even if it is below the ionization threshold; this occurs in particular at the lateral regions of the head. Negative streamers are therefore dissolved easier and conversely they are harder to create; they cannot attain the minimal diameter found in positive streamers, and the field enhancement at their tips is less. This creates the apparently paradoxical situation that negative streamers (if they emerge) are somewhat (about 20%) slower than positive ones though they are supported by the electron drift while positive streamers have to propagate against it. That negative sprites (i.e., downward propagating sprites after a negative cloud-to-ground lightning) are observed so rarely, might be related to the fact that negative streamers are more difficult to start, even though the local field exceeds the ionization threshold.

**3.2 Spectra**

As another input for comparison with sprites, we include new spectral measurements in Fig. 4 for future comparison with the sprite measurements of Kanmae et al. [2007]. As the similarity relations discussed in section 2 hold for arbitrary gases and not just for air, we recently have started investigating sprites on other planets in our laboratory setting; the measurements and results are summarized in an accompanying paper by Dubrovin et al. [2009]. We have used the same method to determine spectra as in that paper, analyzing now discharges in artificial air (a mixture of 80% $N_2$ and 20% $O_2$) at 25 mbar. The spectrum has been acquired with two spectrometers, one for each curve. Two emission systems from neutral molecular nitrogen are indicated, namely the first and second positive systems (FPS and SPS). The 777 nm line from atomic oxygen is present on the flank of one of the FPS bands.

It should be noted that in this discharge, most radiation is produced during a short pulsed glow discharge after the propagation of the streamer heads. However, Nijdam et al. (see the appendix in preprint: http://arxiv.org/abs/0912.0894 - version 1) have shown that the visible spectrum of such a short pulsed glow discharge with pulse durations of 130 ns and with a sufficiently long waiting time until the next pulse is nearly identical to a streamer discharge. The main difference is that the glow discharge has a much higher intensity and therefore leads to a much better signal to noise ratio in the spectrum than the pure streamer discharge.

When one compares spectra of streamers at different pressures one should keep in mind though that the collisional quenching of excited states can break the Townsend scaling (see 2.1.2), similarly to the density effect on photoionization above pressures of 80 mbar. Therefore, the relative intensity of spectral lines in streamer emissions may vary with pressure due to quenching, as suggested recently by Liu et al. [2009]. In experimental measurements in pure nitrogen at pressures from 25 to 200 mbar by Nijdam et al. (http://arxiv.org/abs/0912.0894 - version 1), some changes in relative line intensity are attributed to a decreasing vibrational temperature with decreasing density. Note that sprite spectra measured from low altitudes can also be affected by absorption and scattering in the atmosphere.

## 3.3 Branching and interactions of streamers: mechanisms and 3D structure

*3.3.1 Simulations and theory*

Particular aspects of streamer morphology are branching and the interaction of several streamers. As discussed in section 2.1.4, streamer branching has been simulated, but quantitative predictions for streamers in air still face large methodological challenges, in particular, when going beyond the density approximation. Likewise, only recently the first simulations of interacting streamers could be performed; here two approaches were followed.

First, Luque et al. [2008b] have mastered the numerical problems of fully three-dimensional streamer simulations and presented truly three-dimensional simulations of two negative streamers in air of varying density and in STP mixtures of $N_2$ and $O_2$ with varying mixing ratios. Two streamers extending from the same seed or electrode carry charges of the same polarity in their head; therefore they naturally repel each other electrostatically. However, if the streamers have a strong and long ranged photo-ionization reaction as in air, overlapping ionization clouds between the streamer heads can make them merge, hence overcoming the electrostatic repulsion. As photo-ionization is not quenched at sprite altitudes, the attraction is somewhat stronger than at sea level; this is confirmed by the simulations.

Second, Luque et al. [2008a] analyze a periodic array of strongly interacting streamers. Only Naidis [1996] studied weakly interacting streamers before. Compared to single streamers, that are exclusively studied in all other simulations, interacting streamers do not show the rather homogeneous electric field in their interior, because the charges of neighboring streamer heads contribute to electric screening; in the extreme case considered by Luque et al. [2008a], the electric field in the streamer interior is screened completely at a distance behind the head that is larger than the lateral distance to the neighboring streamers. Furthermore, the closely packed streamers in strong fields can not expand and accelerate as the single ones studied by Arrayas et al. [2002], Rocco et al. [2002], Liu & Pasko [2004, 2006], Montijn et al. [2006b, 2006c], Luque et al. [2007], Liu et al. [2009]. New studies of hexagonal arrays of positive streamers in air in 3D are currently in preparation by Ratushnaya et al.

*3.3.2 Experiments*

Despite these encouraging theoretical results, we suggest that many predictions on branching and interacting sprite streamers can be taken directly from streamer experiments. Briels et al. [2008b] measure the typical length *D* that a streamer propagates before it branches. Positive streamers in air with a diameter *d* propagate a distance $D/d = 11 \pm 4$ before they branch; this ratio is rather independent of *d* if *d* is larger than minimal, and the ratio is constant within the error bar for pressures of 0.1 to 1 bar at room temperature. Nijdam et al. [2008] have largely improved the morphological studies by introducing stereoscopic imaging, resolving the full 3D structure. An explicit result in the 2008 paper is the branching angle; it is approximately Gaussian distributed with $43º \pm 12º$ for pressures from 0.2 to 1 bar. Nijdam et al. [2009] use stereoscopic imaging to analyze apparent reconnection and merging events (where merging is mediated by photoionization while reconnection is due to electrostatic attraction). Indeed cases were found where a streamer channel actually approaches another existing channel of the same discharge at an angle of close to 90º, probably after the other channel has changed polarity. Similar events have been seen earlier in sprites by Cummer et al. [2006], but there it could not be decided whether the connection was real or an artifact of the 2D image projection. The comparison with Nijdam's results suggests that sprite reconnections might be real, and that for understanding such events, one should search for a mechanism where the back end of a sprite channel gains a different polarity than its tip.

**3.4 Streamers as chemical reactors and electron accelerators, X-rays and γ-rays**

In section 2.1.6, we discussed the high mean energies that electrons reach at the tips of growing streamers due to the strong local field enhancement, and we argued that these electrons therefore would essentially not feel the geomagnetic field, in contrast to thermal electrons at the same altitude. The relation between typical electric fields at the streamer head and the mean electron energies is included in Fig. 2. These high electron energies have two other consequences to be elaborated in the following subsections (with references). First, they leave a different distribution of primary molecular excitations behind than, e.g., a stationary glow discharge – this effect is currently being actively explored for various technical applications in plasma chemistry (e.g., for energy efficient destruction of volatile organic components, for air cleaning at highway tunnels or in hospitals, for processing of biogas, or for various disinfection processes, to name but a few). Second, beyond a high energy on average, the electron energy distribution also has a long tail at high energies that is characteristic for the nonequilibrium character of the process; these high-energy electrons can create hard X-ray radiation, with energies certainly exceeding 200 keV [Nguyen et al., 2008].

*3.4.1 Recent experimental results on chemistry and hard radiation from streamers*

Winands et al. [2008] studied streamers in a wire-to-plate electrode geometry and Briels et al. [2008a] in a needle-to-plate geometry. In both cases the streamers were generated by short voltage pulses rising up to 100 kV within a few tens of nanoseconds and with sufficient time lags until the next pulse and discharge. Voltage pulses in the range of 50 to 100 kV in air at standard temperature and pressure (STP) create the fat type of streamers shown in Fig. 1, rather than the much thinner and slower ones at 5 to 30 kV illustrated in Fig. 1 as well. These thick streamers are very interesting sources both of O* radicals and of hard X-rays. While conventional industrial ozone generators work on quite thin streamers, van Heesch et al. [2008] have demonstrated that thick streamers driven by rapidly pulsed voltages of 60 to 100 kV in the wire-to-plate geometry are exceptionally efficient in creating O* radicals and consecutively ozone in air, and that the negative streamers seem to be slightly more efficient than the positive ones. In fact, van Heesch states that more than 50% of the electric energy coupled into his air discharge was converted into ozone.

Nguyen et al. [2010] have shown that positive streamers in the same set-up can emit X-rays with energies between 10 and 42 keV if voltage pulses of 85 kV are applied; this happens during the initiation of the primary streamer near the electrode wire. This demonstrates that a streamer with its high local field enhancement and its local electron energy distribution indeed can accelerate electrons to energies above 42 keV. This is a new step in the recent series of laboratory experiments [Dwyer et al., 2005, 2008; Rahman et al., 2008; Nguyen et al., 2008; Rep'ev & Repin, 2008] aiming to understand X-ray and γ-ray emissions from natural lightning.

*3.4.2 Simulation tools and results on run-away electrons (and X-rays) from streamers*

A 3D simulation of a streamer that follows the motion of all individual electrons would immediately deliver the energies of run-away electrons as well as the distribution of excited molecular levels after the streamer ionization front has passed, but such a simulation does not exist (yet) due to the unmanageable large number of electrons, but methods to work around this limitation just have been developed. The energies of runaway electrons are required for calculating X-ray or even γ-ray emissions, and the primary excitations of molecules are required for calculating the chemical products. The motion of individual electrons is appropriately modeled by a particle model that takes the relevant elastic, inelastic and ionizing collision events between electrons and molecules into account. As mentioned above in section 2.1.6, scattering cross sections for implementation into a particle model are listed, e.g., in the siglo database [Morgan, 1995] and its updates. Below we will only discuss the X-ray aspects of particle modeling.

Whether electrons can run away from a streamer head was first studied by Moss et al. [2006] with a 1D Monte Carlo simulation where the 3D electric field profile was approximated by a stepped function in 1D.

Chanrion & Neubert [2008] developed a 2.5D Monte Carlo model with super-particles. The presently available computing power forced them to present many real electrons by one super-particle. On the one hand, the super-particle approach causes numerical heating and stochastic errors [Li C et al., 2008b], on the other hand, the resolution of the high energy electrons is very low. The second shortcoming of the super-particle approach has now been addressed with an energy-dependent re-sampling of the super-particles by Chanrion & Neubert [2009], which allows them to study the run-away electrons in a negative streamer with much better precision.

Chao Li and coauthors have developed another approach that can efficiently simulate the streamer propagation while the energetic particles are followed with single-particle resolution. Li C et al. [2007] compared results of density and particle models for 1D streamer fronts. Li C et al. [2008a, 2010] coupled density and particle model in 1D, applying the Monte Carlo particle model in the relevant high field region while modeling the many electrons in the streamer interior in an efficient density approximation. Li C et al. [2009] have presented fully 3D hybrid simulations where all single electrons in the high field region of the negative streamer head are followed individually. They find that electrons can gain run-away energies above 200 eV when the field enhancement at the streamer head exceeds 600 Td, which is equivalent to 160 kV/cm at standard temperature and pressure. Chanrion & Neubert [2009] confirm this result, and find independently that a field enhancement of 4.9•32 kV/cm ≈ 160 kV/cm in streamers in STP air can create runaway electrons. (It should be noted that slightly different cross-sections for electron collisions were used, as Li C et al. calculate in pure $N_2$, while Chanrion & Neubert have implemented artificial air ($N_2$:$O_2$ = 80:20), but no photoionization.)

## 4. Summary and conclusions

Streamers are a small but essential part of atmospheric discharges; they play a large role in the early stages of lightning and are physically similar to sprites. In the present paper, we have reviewed those recent experiments and also simulations of streamers that are applicable to processes of atmospheric electricity, and we have discussed how to compare them to sprites.

In the introduction, we reviewed present theoretical approaches that nowadays are able to include processes on different length scales, but never the whole range from the motion of individual accelerated electrons at the streamer tip up to the ground-ionosphere distance. We then suggested considering laboratory experiments of streamers as simulations of sprites.

To create a basis for such a comparison, section 2 discusses first the theoretical aspects of the similarity of streamers and sprites, and then the comparison of experiments or observations. The similarity of streamer discharges at different gas density is essentially based on the fact, that ionization energies are independent of density, while length scales are set by the mean free path length of the electron and scale with inverse density. Corrections to the similarity relations come (a) from processes in the streamer channel, (b) from electrodes or other material boundaries that do not vary with pressure, (c) from different statistical fluctuations, as streamers at lower densities contain more electrons, (d) from external ionization sources or spatial changes of gas density, (e) from (geo-)magnetic fields or (f) from gas heating.

For the last two cases, we present new results: We argue that the geomagnetic field at mesospheric altitudes certainly has an effect on thermal electrons, but not on the very energetic electrons at the streamer tip. Furthermore, we argue that the gas in a streamer channel heats up easier at higher densities, because the relative electron density is higher. We therefore suggest that a sprite streamer in the mesosphere is unlikely to create so much Ohmic heat that it can transit into a hot leader, while on the opposite a streamer at liquid densities will heat up so rapidly that it might transit directly into a hot leader phase.

The physical similarity of experimental streamers and sprites is confirmed by the following observations. The morphology is similar, the measured minimal diameters are related by similarity relations, and the minimal velocities are the same. The light emission is focused at the growing tips both in streamers and in sprites. Finally experimental investigations of streamers in high magnetic fields are consistent with observations of sprites in the geomagnetic field.

In section 3, a number of measurements of streamers are discussed and related to sprites, and occasionally also sprite relevant streamer theory is included into the discussion. First the morphology of streamer trees and the large range of streamer diameters and velocities and their electric currents are discussed. Typically, in a needle-to-plane electrode geometry a high and fast voltage pulse generates thick and fast streamers; these streamers branch into thinner and slower streamers, those branch again, until the thinnest and slowest streamers emerge. These streamers of minimal width and velocity do not branch anymore, but extinguish after some propagation distance. We suggest that sprites have a similar variety of diameters and velocities, but that the sprites of minimal diameter are generically difficult to detect while streamers of minimal diameter are the easiest to make. A sprite streamer that branches is not minimal.

The section contains new measurements of spectra of streamers in air that parallel our recent investigation of sprites, their structure and their spectra on Venus and Jupiter [Dubrovin et al., 2009]; as the similarity relations do not refer to any specific gas type, they are applicable in other gas compositions as well.

Next the streamer tree morphology is analyzed in more detail. Two types of streamer-streamer interaction are identified, namely through electrostatic forces or through nonlocal photo-ionization. Then recent experimental results are reviewed, namely propagation length until branching, distribution of branching angles, and true or fake interactions of channels; the last two results are based on stereoscopic imaging and 3D reconstruction.

Finally, the chemical and radiation products of streamer discharges are reviewed. These concern experimental results on the high chemical efficiency of thick streamers and on their energetic radiation, and recent theory on streamers as sources of run-away electrons.

The authors thank H. Stenbaek-Nielsen and an anonymous referee for very valuable suggestions. They acknowledge financial support by STW-projects 06501 and 10118 of the Netherlands' Organization for Scientific Research NWO, by the Dutch National program BSIK, in the ICT project BRICKS, theme MSV1, and by the Dutch IOP-EMVT under contract number 062126B.

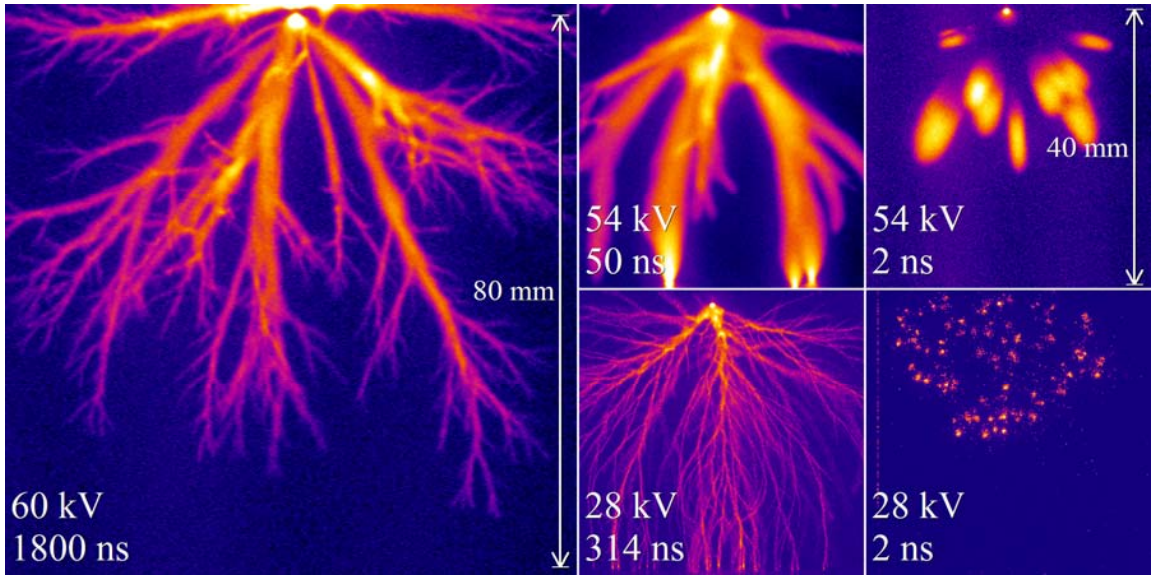

**Figure 1**: Photographs of positive streamers in air at standard pressure and temperature scaled to identical lengths. 80 or 40 mm indicate the distance from the electrode needle above to the electrode plate below. Applied voltage and exposure time of the camera are given in the lower left corners of the panels. Essentially the same figures and an explanation of the experiments can be found in [Briels et al., 2006, 2008a; Ebert et al., 2006]. The 80 mm and the 40 mm discharges driven by ~60 kV have similar electric fields near the needle electrode and initially produce similarly thick streamers that branch into thinner streamers. The thin streamers half way in the 80 mm gap resemble very much the thin streamers in the 40 mm gap powered by 28 kV. From these pictures as well as from theoretical considerations the conclusion can be drawn that a thick and well conducting streamer can generate a similar streamer corona as a needle electrode. The voltage at the tip of such a "streamer needle" is lower than at the generating electrode due to the decrease of the electric field along the length of the streamer.

The middle and the right column show the difference between an exposure time long enough for the streamers to cross the gap and a short exposure time of 2 ns. The glowing dots marking the heads of the growing streamers with the 2 ns exposure are much larger for 54 kV. In fact, as detailed by Briels et al. [2006, 2008a], the streamer diameter increases by a factor of 6 and the velocity by a factor of 15, when the voltage is increased by less than a factor of 2 from 28 to 54 kV.

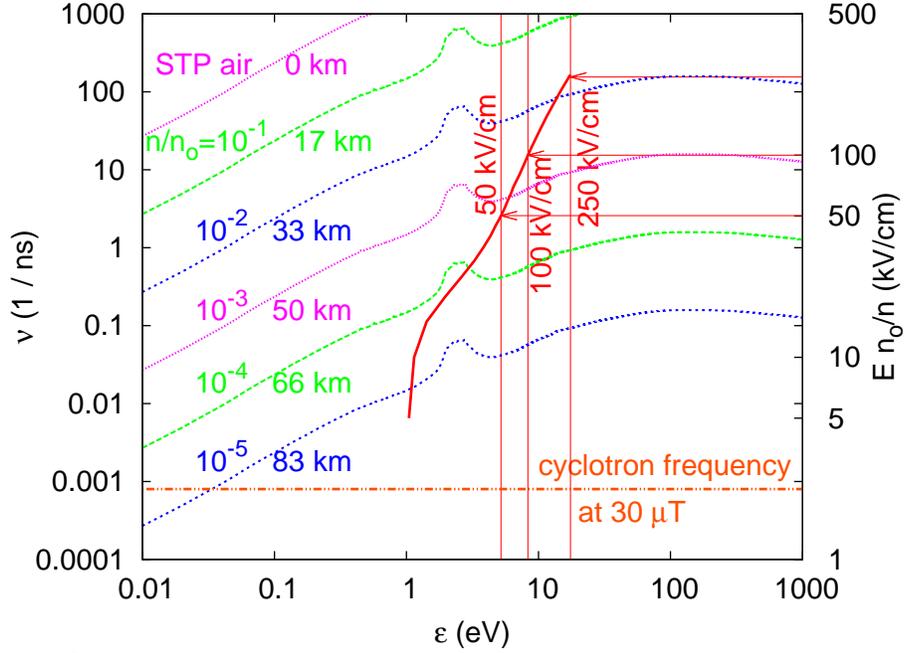

**Figure 2:** The influence of the geomagnetic field as a function of air density and mean electron energy in the discharge, with the relation between mean electron energy $\epsilon$ (bottom axis) and reduced electric field (right axis). Plotted are the collision frequencies $\nu$ (left axis) as a function of the mean electron energy (bottom axis) for electrons in air from ground pressure (STP air) up to 83 km altitude as dotted lines (with air density approximated as $n = n\_0 \exp[-z/h]$, h=7.2 km). The collision frequencies are determined from 44 electron-neutral collision processes for an air mixture of 78.084 % N2, 20.982 % O2, and 0.934 % Ar taken from the Siglo database [Morgan, 1995]. The dash-dotted line indicates the cyclotron frequency $\nu_c = qB/(2\pi m) = 0.8$ MHz at the geomagnetic field of 30 μTesla close to the equator. The solid red line represents the relation between the reduced electric field (right axis) and the mean energies of an electron swarm in this field (bottom axis), calculated as in [Li C et al., 2007]. To help readers, characteristic reduced fields in the streamer head of 50, 100 and 250 kV/cm are indicated with a thin horizontal arrow and the corresponding mean electron energies (of 5.2, 8.3, and 17.5 eV) with a thin vertical line. It can be seen that the electron collision frequency is much larger than the cyclotron frequency for all these electron energies, even at 83 km altitude; this means that the effect of the geomagnetic field on streamers and sprites is small. We recall that mean electron energy of 1 eV corresponds to a temperature of 11 600 K, or 0.01 eV to 116 K.

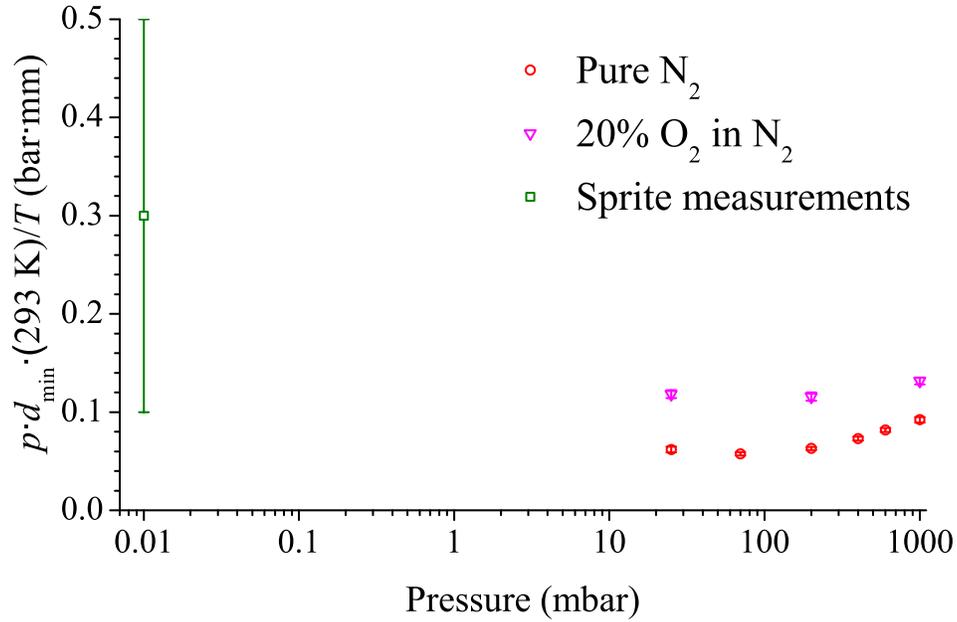

**Figure 3:** Reduced minimal streamer diameter $p \cdot d_{min}\, 293K/T \sim n \cdot d_{min}$ (according to the ideal gas law $p = n\, k_B T$) as a function of pressure $p$; here T is temperature and 293 K is room temperature, and the diameter is determined as the full width at half maximum of the light emission. Triangles: experimental results in artificial air (a mixture of 20% $O_2$ in $N_2$). Circles: experimental results in pure $N_2$. Square: minimal sprite diameter at 80 km altitude from Gerken et al. [2000], evaluated as discussed by Briels et al. [2008a]; the error bars . The laboratory results are from Nijdam et al. [2009b] and are an improvement of the measurements presented by Briels et al. [2008a]. The reduced minimal streamer diameter at room temperature is here found to be ~ 0.12 mm bar.

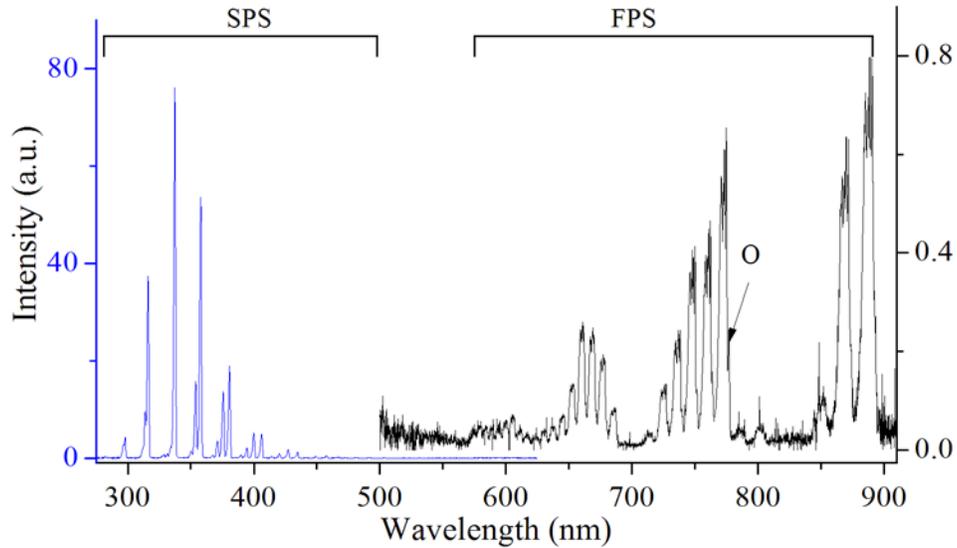

**Figure 4:** Spectra of streamers and of the consecutive short pulsed glow discharges, measured with two spectrometers of different spectral range. Each spectrometer has delivered one curve. The curves have been corrected for the sensitivity of the spectrometers and their scales are comparable. The spectrum is dominated by the Second Positive System of molecular nitrogen. This system is about 100 times stronger than the First Positive System, also of molecular nitrogen. Besides these two systems, the only other significant feature is an oxygen line at 777 nm. This spectrum is discussed in more detail by Nijdam et al. (see the appendix in preprint: http://arxiv.org/abs/0912.0894 - version1).